\documentclass[prb,letterpaper,aps,floatfix,twocolumn,superscriptaddress]{revtex4-2}

\usepackage{graphicx}
\usepackage{amsmath,amssymb}
\usepackage{bm}

\usepackage[pdfstartview=FitH,breaklinks=true,bookmarks=true,colorlinks=true,anchorcolor=black,citecolor=red,filecolor=black,menucolor=black,urlcolor=blue,linkcolor=blue]{hyperref}

\begin{document}

\title{Detecting quantum phase transitions via shallow variational quantum circuits}

\author{Ching-Yu Huang}
\email{cyhuangphy@thu.edu.tw}
\affiliation{Department of Applied Physics, Tunghai University, Taichung 40704, Taiwan}
\affiliation{Physics Division, National Center for Theoretical Sciences, Taipei 10617, Taiwan}
\affiliation{Center for Theoretical and Computational Physics, National Yang Ming Chiao Tung University, Hsinchu 300093, Taiwan}
\author{Min-Fong Yang}
\email{mfyang@thu.edu.tw}
\affiliation{Department of Applied Physics, Tunghai University, Taichung 40704, Taiwan}

\date{\today}

\begin{abstract}
Mapping quantum phase diagrams through classical simulation is notoriously resource-intensive, as even small systems far from the thermodynamic limit demand prohibitive computational effort. The variational quantum eigensolver (VQE) offers a compelling alternative, exploiting approximate ground states to distinguish phases. An appealing proposal, dubbed as Delta-VQE, determines critical points by contrasting variational energies optimized from reference states of distinct phases. Intriguingly, the diagnostic sharpens as circuit depth decreases, highlighting its promise as a resource-conscious probe of quantum criticality.
To probe the broader applicability and underlying mechanisms of this approach, we investigate the one-dimensional transverse-field Ising model with a three-spin cluster interaction, a setting in which the Ising transitions are generally situated beyond the self-dual line. We demonstrate that, whenever dual ansatze are employed, Delta-VQE invariably detects the self-dual points rather than the true criticality. In contrast, when ansatze are carefully tailored to embody the competing phases across the boundary, the genuine Ising critical point can be successfully identified with only minor finite-size effects. Our results establish that, while Delta-VQE provides a resource-efficient probe of quantum criticality without requiring precise ground-state preparation, its diagnostic power is fundamentally contingent upon the judicious selection of physically representative ansatze.
\end{abstract}

\maketitle

\section{Introduction}

Understanding quantum phases of matter and the transitions among them is a central theme in quantum many-body physics. In computational condensed matter studies, such transitions are typically investigated by determining the ground state of the system and measuring the corresponding order parameters. Distinct phases are commonly characterized through appropriate order parameters or other physical observables. However, these calculations are often challenging or intractable within classical simulations, as they demand substantial computational resources even for relatively small systems.

Recent advances in quantum computing have opened new possibilities for studying quantum many-body systems. By exploiting superposition and the natural encoding of entanglement,
quantum computers can encode exponentially large many-body quantum states in quantum registers, thereby probing properties beyond the reach of classical methods. Among the most promising applications of quantum computing is the preparation of ground states of interacting quantum Hamiltonians and the subsequent exploration of quantum phase transitions therein.

Owing to the limited number of qubits and imperfect gate fidelities,
current quantum hardware is considered to be in the noisy intermediate-scale quantum (NISQ) regime~\cite{Brooks_2019}. A class of algorithms particularly well-suited to NISQ conditions is the variational quantum eigensolver (VQE)~\cite{Peruzzo_etal2014,Cerezo_etal2021,Tilly_etal2022}, which aims to prepare the ground state of a quantum Hamiltonian via the variational principle.
VQE is a hybrid quantum–classical algorithm in which parameterized quantum circuits prepare trial states, while classical optimization routines iteratively adjust the circuit parameters to best approximate the solution. By bridging quantum resources with classical efficiency, VQE offers a promising alternative route that may prove advantageous for certain strongly correlated systems. Nonetheless, VQE optimization may be hindered by barren plateaus and local minima~\cite{Larocca_etal2025}. These challenges can sometimes be addressed through tailored ansatz design or by focusing on noise-robust observables.

In recent years, the application of VQE to quantum phase transitions has attracted considerable attention~\cite{Bravo-Prieto_etal2020,Dreyer_etal2021,Jobst_etal2022,%
Schneider_etal2026,Shi_etal2023,Lively_etal2024,Bosse_etal2024,Cao_etal2025,%
Crognaletti_etal2025,Duriez_etal2025,Dev-Sharma2025}.
This growing body of work presents a diverse set of methodologies for locating critical points and characterizing phase boundaries in many-body models. A key advantage in this context is that faithfully preparing the exact ground state is often unnecessary for constructing phase diagrams; low-fidelity approximations may suffice to reveal qualitative features and locate critical points.
Existing approaches employ a variety of observables.
Some monitor conventional local order parameters~\cite{Dreyer_etal2021,Jobst_etal2022}, whereas others identify phase transitions through derivatives of the variational ground-state energy with respect to Hamiltonian parameters~\cite{Lively_etal2024,Duriez_etal2025,Dev-Sharma2025}  or circuit depth~\cite{Bosse_etal2024}.
Beyond conventional observables, machine-learning techniques have also been incorporated into VQE. In cases where traditional order parameters are unknown or insufficient, particularly for topological phase transitions, VQE has been integrated with classical machine learning to identify transitions by examining locally trapped states in the optimization landscape~\cite{Cao_etal2025}.
Besides, inspired by the level spectroscopy method~\cite{Okamoto-Nomura1992,Nomura-Okamoto1994,Nomura1995,Nomura-Kitazawa1998,%
Nomura-Kitazawa2002}, Crognaletti~\emph{et al}.~\cite{Crognaletti_etal2025}
advanced an equivariant VQE that embeds physical symmetries into the targeted low-lying excited states. By monitoring energy-level crossings among these states, the Berezinskii–Kosterlitz–Thouless transition~\cite{BKT_book} in the $J_1-J_2$ Heisenberg chain is precisely located, where direct order-parameter measurements are hindered by finite-size effects.
Collectively,  these studies show that VQE is not merely a tool for ground-state energy estimation; it constitutes a versatile framework for probing complex quantum critical phenomena through circuit design, symmetry preservation, and sophisticated post-processing of variational outputs. By enabling the exploration of systems and parameter regimes that remain challenging for classical methods, VQE has established itself as a powerful and adaptable approach to investigating quantum criticality.

Moreover, Shi \emph{et al.}~\cite{Shi_etal2023} proposed a novel approach termed Delta-VQE.
Instead of focusing on a single ground state, it compares the variational energies ($E_1$ and $E_2$) obtained by initializing the system in two distinct reference states that correspond to different phases of matter. The critical point is then identified as the parameter value at which the absolute energy difference, $\Delta E = |E_1 - E_2|$, is minimized. Interestingly, this signature becomes more pronounced with shallower quantum circuits, thereby providing a resource-efficient probe of criticality. In short, Delta-VQE embodies a conceptual shift in the use of VQE for phase transition detection, emphasizing the qualitative contrast between phases over the quantitative exactness of ground-state energy evaluation. This algorithm has been validated across several quantum models, including the transverse-field Ising model, the spin-anisotropic XZ model, the clustering Ising model, and the XY-Ising model. Notably, the exact critical points are reproduced even with a single-layer ansatz and for small system sizes in the first three models.
The success of this approach highlights the robustness of $\Delta E = |E_1 - E_2|$ as a diagnostic, even when the individual variational energies are themselves inaccurate due to the shallow circuit depth employed. Nevertheless, as remarked by the authors in Ref.~\cite{Shi_etal2023}, Delta-VQE is a physics-inspired heuristic method, and a rigorous proof of its broader applicability is still lacking. This observation motivates the development of a more general criterion for understanding and extending the applicability of Delta-VQE.

In this work, we aim to explore the general validity of Delta-VQE and to uncover the underlying principles governing its effectiveness. To this end, we consider the transverse-field Ising model with a three-spin cluster interaction [see Eq.~\eqref{eq:Hami}]~\cite{Kopp-Chakravarty2005,Smacchia_etal2011, Montes-Hamma2012,Niu_etal2012,Ding2019,Yu-Li2024}. This model exhibits self-duality when the transverse field $h=1$, independent of the Ising coupling $J$. However, the actual Ising transitions occur at $h_c=1\pm J$~\cite{Kopp-Chakravarty2005}. We find that, when dual variational ansatze are employed, $\Delta E=0$ consistently arises at $h=1$, even though this point deviates from the true critical points. By contrast, when appropriate initial states are chosen, the signal of the genuine Ising transitions can be revealed. This observation underscores that the effectiveness of Delta-VQE is highly contingent upon selecting representative initial states for the distinct phases. As with most numerical approaches, the estimated critical points for systems of small sizes do not precisely coincide with the exact values. Nevertheless, the deviations can be considerably smaller than those obtained by certain conventional methods (e.g., the fidelity approach; see Sec.~\ref{sec:results}). Moreover, when the system size $N$ and circuit depth $p$ are increased simultaneously according to the finite-depth scaling relation $N \approx 4p$, the agreement between the Delta-VQE estimates and the exact results is substantially improved.

In addition, we investigate an extended transverse-field Ising model in App.~\ref{appendix:eTFIM} to further support the general applicability of the Delta-VQE approach in identifying quantum critical points. Furthermore, while the main text focuses on ideal noiseless simulations, the impact of realistic quantum-device noise on the Delta-VQE protocol is examined in App.~\ref{appendix:noise}. We show that the noiseless results of the main text can be faithfully recovered through an appropriate zero-noise extrapolation procedure. This finding demonstrates the resilience of Delta-VQE against noise and suggests its potential applicability in the NISQ regime.

The remainder of this paper is organized as follows.
We introduce our model and outline the framework of Delta-VQE in Sec.~\ref{sec:method}.
In Sec.~\ref{sec:results}, we present our Delta-VQE results to illustrate both its performance and its limitations.
We conclude our paper in Sec.~\ref{sec:conclusion}.

\section{model Hamiltonian and Delta-VQE}\label{sec:method}

In this section, we begin by introducing the model under consideration and describing its possible phases and phase boundaries. We then present the detailed procedures of Delta-VQE as applied to this model.

\subsection{generalized cluster-Ising model}\label{sec:model}

The Hamiltonian $H$ of the one-dimensional transverse-field Ising model with a three-spin cluster interaction (also referred to as the generalized cluster-Ising model)~\cite{Kopp-Chakravarty2005,Smacchia_etal2011,%
Montes-Hamma2012,Niu_etal2012,Ding2019,Yu-Li2024} is given by
\begin{equation}\label{eq:Hami}
H=-J \sum_{i} X_{i} X_{i+1} -h \sum_{i} Z_{i} -\sum_{i} X_{i-1} Z_{i} X_{i+1}
\end{equation}
with $X_{i}$ and $Z_{i}$ being the Pauli matrices at site $i$.
Here, the coupling strength of the three-spin cluster interaction is set to unity, and periodic boundary conditions are assumed. This model possesses a global $\mathbb{Z}_2$ symmetry generated by $\prod_i Z_i$, under which $X_i\rightarrow -X_i$. However, a symmetry-breaking transition occurs when the Ising coupling $J$ becomes sufficiently strong. As a result, the system exhibits four distinct phases. The system is in the paramagnetic (PM) phase when the transverse field dominates. In this phase, the spins tend to align along the field direction, the $\mathbb{Z}_2$ symmetry is preserved, and the excitation spectrum is gapped. When the three-body cluster interaction is dominant, the system enters a symmetry-protected topological (SPT) cluster phase characterized by a non-zero string order parameter. When the nearest-neighbor Ising interaction prevails, the system undergoes $\mathbb{Z}_2$ symmetry breaking and becomes ferromagnetic (FM) for $J>0$ or antiferromagnetic (AFM) for $J<0$. In our study, we focus on the regime $J>0$, as the analysis for $J<0$ yields qualitatively similar conclusions.

Similar to the conventional transverse-field Ising model with Kramers–Wannier duality, this generalized cluster-Ising model likewise exhibits exact self-duality. By defining a new cluster duality mapping $\tilde{Z}_i = X_{i-1} Z_i X_{i+1}$ and $\tilde{X}_i = X_i$, the transformation swaps the one-body transverse field term with the three-body cluster interaction term, while leaving the Ising coupling $X_i X_{i+1}$ unchanged. This duality transformation corresponds precisely to the parameter exchange $h \leftrightarrow 1$. Consequently, the condition $h=1$ defines the self-dual line of the system.

\begin{figure}[t]
\includegraphics[width=\linewidth]{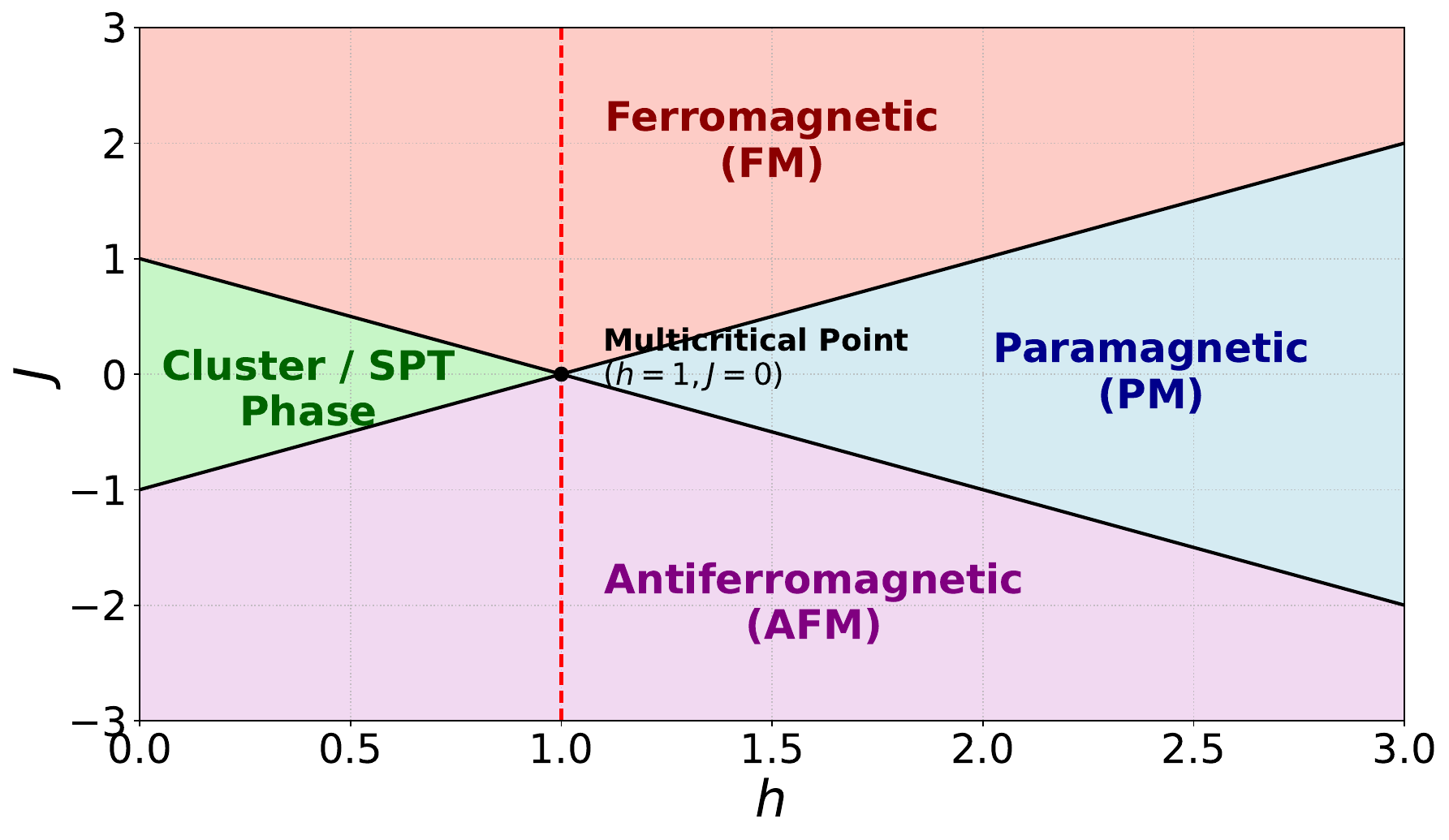}
\caption{Phase diagram of the model in Eq.~\eqref{eq:Hami}. The black solid lines  denote the phase boundaries of Ising transitions, $h_c=1\pm J$, which cross at the multicritical point $(h, J)=(1, 0)$. The red dashed line indicates the self-dual line. }\label{fig:phase_dia}
\end{figure}

However, unlike the conventional transverse-field Ising model, the self-dual points of the present system do not always coincide with the critical points. This correspondence occurs only when $J=0$. The distinction can be understood by mapping the spin model into a free-fermion representation via the Jordan–Wigner transformation. In this formulation, the excitation gap is found to close at $h_c=1\pm J$~\cite{Kopp-Chakravarty2005}, which defines the true phase boundaries of the Ising transitions. These two critical lines partition the entire $h-J$ parameter space into four distinct physical phases, as depicted in Fig.~\ref{fig:phase_dia} (see also Fig.~2 of Ref.~\cite{Kopp-Chakravarty2005}). In this phase diagram, the self-dual line corresponds to the vertical line $h=1$. As discussed above, except for the single point at $J=0$, the entire dual line $h=1$ lies within the interior of gapped ordered phases. It is therefore not a phase boundary, but rather a dual symmetry line embedded within the gapped phases.

\subsection{framework of Delta-VQE}\label{sec:Delta-VQE}

While quantum critical points can, in principle, be located by accurately solving the ground state, the quantum resource requirements are prohibitive for NISQ devices.
To address this challenge, Z.-Q. Shi et al.~\cite{Shi_etal2023} proposed the Delta-VQE algorithm, which leverages the intrinsic asymmetry of shallow quantum circuits across phase boundaries.

The algorithmic procedure of Delta-VQE is defined as follows:
\begin{enumerate}
    \item Starting from two distinct reference states, $|\psi_1\rangle$ and $|\psi_2\rangle$, corresponding to two different quantum phases of matter, we construct two families of parameterized trial states:
    \begin{align}
        |\psi_1(\bm{\theta}_1)\rangle &= U_1(\bm{\theta}_1)|\psi_1\rangle \; , \nonumber \\
        |\psi_2(\bm{\theta}_2)\rangle &= U_2(\bm{\theta}_2)|\psi_2\rangle \; . \nonumber
    \end{align}
    Here, $U_i(\bm{\theta}_i)$ ($i=1,2$) denotes a unitary operator parameterized by a set of circuit angles $\bm{\theta}_i$. Note that the unitary ansatze of $U_1$ and $U_2$ need not share the same structure in order to ensure non-trivial optimization. When the ansatze are sufficiently expressive, approximate ground states can be reached for suitable choices of the parameters $\bm{\theta}_i$.

    \item The optimized variational parameters $\bm{\theta}_i^*$ are obtained by minimizing the energy functional $E_i(\bm{\theta}_i;h,J) = \langle\psi_i(\bm{\theta}_i)|H(h,J)|\psi_i(\bm{\theta}_i)\rangle$ using classical optimization routines. The classical optimizer iteratively updated these parameters until the specified convergence criterion or the maximum number of iterations was reached. The optimized variational energies are then obtained respectively as:
    \begin{align}
        \begin{aligned}
            E_1(h,J) &= \langle\psi_1(\bm{\theta}_1^*)|H(h,J)|\psi_1(\bm{\theta}_1^*)\rangle \; , \\
            E_2(h,J) &= \langle\psi_2(\bm{\theta}_2^*)|H(h,J)|\psi_2(\bm{\theta}_2^*)\rangle \; .
        \end{aligned}\label{eq:opt_E}
    \end{align}

    \item Calculate the absolute energy difference functional:
    \begin{equation}\label{eq:delta_E}
        \Delta E(h,J) = |E_1(h,J) - E_2(h,J)| \; .
    \end{equation}
    The minimum point of $\Delta E(h,J)$ for a given $J$ is identified as the quantum critical point $h_c$.
\end{enumerate}

Within the Delta-VQE protocol, the reference states are chosen to represent distinct quantum phases, and the Hamiltonian variational ansatz (HVA)~\cite{Wecker_etal2015,Ho-Hsieh2019,Wiersema_etal2020} is employed to construct the parameterized quantum circuits. The HVA is designed to construct a sequence of quantum gate operations tailored to the specific Hamiltonian. The first step in building an HVA circuit is to decompose the Hamiltonian into a set of non-commuting terms, after which the ansatz is realized through successive applications of sub-Hamiltonian evolutions. The number of sequential operations defines the circuit depth, which directly influences the expressivity and entanglement capability of the circuit.
Moreover, the ansatz inherently respects physical symmetries because HVA operators commute with the system's symmetry operators. If initialized in the target symmetry subspace, the optimization remains strictly confined to that sector, filtering out non-physical excitations. Explicitly preserving these symmetries drastically reduces the variational search space, allowing HVA to exhibit superior optimization properties compared to generic, highly expressive ansatze~\cite{Wiersema_etal2020}.

The physical mechanism underlying the success of this approach can be understood from the convergence properties of variational quantum algorithms across different quantum phases. As established in Ref.~\cite{Dreyer_etal2021} under the condition of system size $N=4p$, the convergence of the variational energy toward its exact thermodynamic-limit value depends crucially on whether the initial reference state and the target state belong to the same quantum phase. This gives rise to distinct scaling behaviors with respect to the circuit depth $p$. When the initial and target states reside within the same gapped phase, the variational energy converges exponentially fast to its thermodynamic-limit value. In contrast, when the target state can only be reached by traversing a quantum phase transition, the symmetry-preserving nature of the variational circuit causes it to retain memory of the initial configuration, resulting in a slow algebraic convergence of $\mathcal{O}(p^{-1})$. Remarkably, at the quantum critical point, the energy error is governed by universal conformal-field-theory data and exhibits a faster algebraic decay of $\mathcal{O}(p^{-2})$. Although these conclusions are obtained from the exactly solvable one-dimensional transverse-field Ising model, they are expected to remain qualitatively valid for other systems exhibiting Ising-type phase transitions.

Translating these observations to the present Delta-VQE study provides a natural explanation for the effectiveness of the method. Away from criticality ($h\neq h_c$), at least one of the two variational states must approximate a target state belonging to a different phase than its reference state, leading to a relatively slow convergence of the energy difference $\Delta E$. By contrast, at the critical point ($h=h_c$), both variational states are expected to exhibit the faster finite-depth scaling suggested by Ref.~\cite{Dreyer_etal2021}. Consequently, $\Delta E$ is suppressed more efficiently in the critical regime and is therefore expected to attain its minimum in the vicinity of the critical point. This provides a theoretical rationale for the Delta-VQE criterion that identifies quantum criticality through the minimum of $\Delta E$. More importantly, as found below, this mechanism remains effective even for relatively shallow circuits, allowing reasonably accurate estimates of critical points with only modest quantum resources.

\subsection{construction for present model}\label{sec:construction}

Having outlined the general procedure, we now turn to the details of our construction. For convenience, the Hamiltonian in Eq.~\eqref{eq:Hami} is rewritten as a sum of non-commuting terms, $H(h,J) = -J H_{XX} - h H_Z - H_{XZX}$, where $H_{XX}=\sum_i X_i X_{i+1}$, $H_Z=\sum_i Z_i$, and $H_{XZX}=\sum_i X_{i-1} Z_i X_{i+1}$. These terms represent competing interactions and quantum fluctuations in the system.
For the FM and PM phases, the reference states are chosen as ground states of $-H_{XX}$ and $-H_Z$, respectively:
\begin{equation}\label{eq:ref_states}
|\psi_\mathrm{F}\rangle = |\mathrm{GHZ}\rangle, \qquad %
|\psi_\mathrm{P}\rangle = |0\rangle^{\otimes N} \; .
\end{equation}
Here, $|\mathrm{GHZ}\rangle \equiv \frac{1}{\sqrt{2}} \left(|+\rangle^{\otimes N}+|-\rangle^{\otimes N}\right)$ denotes the $N$-qubit Greenberger--Horne--Zeilinger (GHZ) state in the $x$ basis. Because spontaneous symmetry breaking does not occur in a finite system, we use the symmetry-preserving GHZ state as the reference for the FM phase, rather than either of the symmetry-breaking product states $|+\rangle^{\otimes N}$ and $|-\rangle^{\otimes N}$. This choice yields improved quantitative accuracy.
For the cluster phase, the reference state is taken to be the cluster state, which corresponds to the ground state of $-H_{XZX}$:
\begin{equation}
|\psi_\mathrm{C}\rangle = \mathcal{H}^{\otimes N}\left(\prod_i CZ_{i,i+1}\right)|+\rangle^{\otimes N} \; .
\end{equation}
Here $\mathcal{H}$ denotes the Hadamard gate and $CZ_{i,i+1}$ is the controlled-Z (CZ) gate acting on qubits $i$ and $i+1$.

The HVA is constructed through successive applications of sub-Hamiltonian evolutions $\exp(i\theta_\alpha H_\alpha)$, where $H_\alpha\in\{H_{XX},H_Z,H_{XZX}\}$ and $\theta_\alpha$ denotes the corresponding variational parameter. By construction, the multilayer unitary operator $U_i(\bm{\theta}_i)$ preserves the symmetries of the target Hamiltonian. Since each sub-Hamiltonian evolution leaves its corresponding reference state invariant, distinct operator orderings are required for different reference states to guarantee non-trivial optimization. In the present study, we employ the following ansatze for the reference states $|\psi_\mathrm{F}\rangle$, $|\psi_\mathrm{P}\rangle$, and $|\psi_\mathrm{C}\rangle$:
\begin{align}
&|\psi_\mathrm{F}(\bm{\theta})\rangle = \prod_{\ell=1}^{p} e^{i\,\theta_3^{(\ell)} H_{XX}}\, e^{i\,\theta_2^{(\ell)} H_Z}\, e^{i\,\theta_1^{(\ell)} H_{XZX}}\, |\psi_\mathrm{F}\rangle \;, \label{eq:psi_F} \\
&|\psi_\mathrm{P}(\bm{\theta})\rangle = \prod_{\ell=1}^{p} e^{i\,\theta_3^{(\ell)} H_Z}\, e^{i\,\theta_2^{(\ell)} H_{XZX}}\, e^{i\,\theta_1^{(\ell)} H_{XX}}\, |\psi_\mathrm{P}\rangle \;, \label{eq:psi_P} \\
&|\psi_\mathrm{C}(\bm{\theta})\rangle = \prod_{\ell=1}^{p} e^{i\,\theta_3^{(\ell)} H_{XZX}}\, e^{i\,\theta_2^{(\ell)} H_Z}\, e^{i\,\theta_1^{(\ell)} H_{XX}}\, |\psi_\mathrm{C}\rangle \;. \label{eq:psi_C}
\end{align}
Here, $p$ denotes the number of HVA layers and therefore controls the circuit depth. These operator orderings avoid trivial initial evolutions and ensure that each trial state contains $3p$ nonredundant variational parameters. Because the HVA evolution strictly preserves symmetry, all these ansatze respect the $\mathbb{Z}_2$ symmetry of the Hamiltonian in Eq.~\eqref{eq:Hami}. We note that $|\psi_\mathrm{P}(\bm{\theta})\rangle$ and $|\psi_\mathrm{C}(\bm{\theta})\rangle$, introduced in Eqs.~\eqref{eq:psi_P} and \eqref{eq:psi_C}, form a dual pair. This duality stems from the dual transformation that exchanges $H_{XZX}$ with $H_Z$ and swaps the states $|\psi_\mathrm{P}\rangle$ and $|\psi_\mathrm{C}\rangle$, while leaving $H_{XX}$ invariant, as discussed in Sec.~\ref{sec:model}.

In what follows, we apply Delta-VQE at a fixed shallow circuit depth $p$ and select the appropriate pairs of ansatze to locate the transitions between the corresponding quantum phases.

\section{numerical results}\label{sec:results}

In our Delta-VQE calculations, we employed the Qiskit software development kit (SDK)~\cite{QISKIT} to construct the parameterized quantum circuits and evaluate expectation values. All simulations were performed using an ideal statevector simulator without noise.
The variational parameters were optimized with the limited-memory Broyden–Fletcher–Goldfarb–Shanno algorithm with bound constraints (L-BFGS-B)~\cite{L-BFGS-B}, as implemented in the SciPy optimization library~\cite{SCIPY}. L-BFGS-B is a gradient-based quasi-Newton method that approximates the inverse Hessian using information from a limited number of previous iterations, thereby reducing memory requirements and making it well-suited for optimizing high-dimensional variational parameter spaces.

Because the variational energy landscape may contain multiple local minima, we implemented a multi-start optimization scheme. For each parameter point, we used the optimized parameters from the preceding point as the initial guess, generated several additional starting points by adding small random perturbations, and performed independent optimizations from each.
The solution with the lowest variational energy was retained. This strategy improves the continuity of the optimized variational solution across neighboring parameter points while reducing the likelihood of convergence to local minima. As outlined in Sec.~\ref{sec:Delta-VQE}, the cost function was defined as the energy functional, and the optimized variational energies were selected according to Eq.~\eqref{eq:opt_E} by retaining the lowest-energy solution among all independent runs.

While the main text focuses on ideal noiseless Delta-VQE simulations, the effects of realistic quantum-device noise on the protocol are investigated in App.~\ref{appendix:noise}. We show that, with an appropriate zero-noise extrapolation procedure, the noiseless results can be faithfully recovered. These findings demonstrate the resilience of the Delta-VQE approach against noise and support its practical applicability on near-term quantum devices.

\subsection{Fidelity susceptibility as a benchmark}

As a benchmark for our Delta-VQE results, we first identify the quantum critical points using the ground-state fidelity susceptibility obtained from exact diagonalization (ED), which is a well-established probe of quantum phase transitions (see Ref.~\cite{fidelity_review} for a review).

For two normalized many-body ground states corresponding to nearby parameter values, $|\Psi_G(h)\rangle$ and $|\Psi_G(h+\delta h)\rangle$, the ground-state fidelity is defined as
\begin{equation}
F(h,h+\delta h) =
\left| \langle \Psi_G(h) | \Psi_G(h+\delta h) \rangle \right|,
\end{equation}
and the corresponding fidelity susceptibility is given by
\begin{equation}
\chi_F(h) = -\lim_{\delta h\rightarrow0} \frac{2\ln F(h,h+\delta h)} {(\delta h)^2}.
\label{eq:FS}
\end{equation}

\begin{figure}[t]
\includegraphics[width=0.9\linewidth]{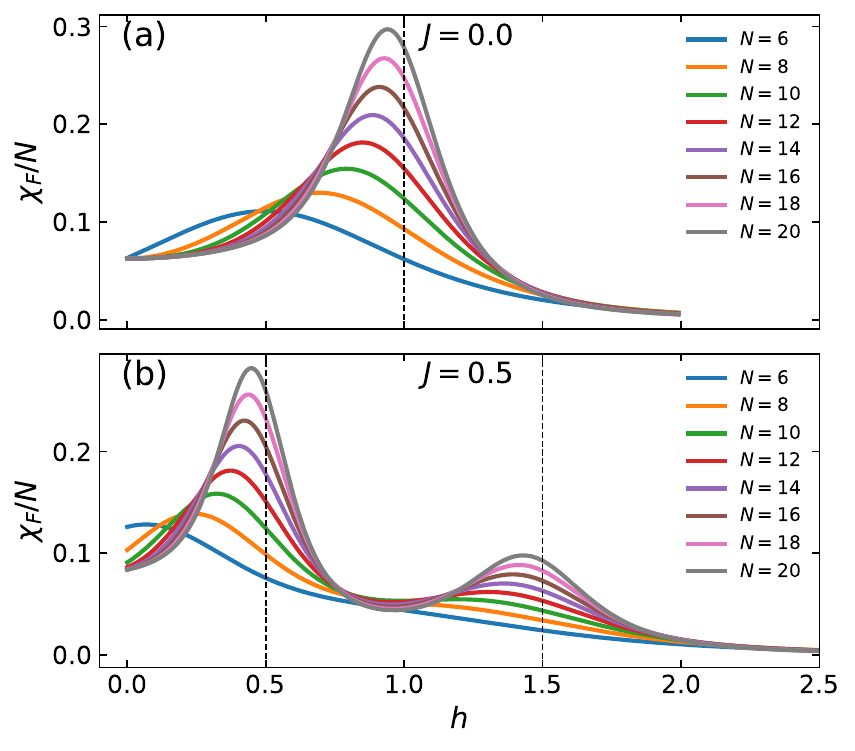}
\label{fig:FS}
\caption{Fidelity susceptibilities per site $\chi_F/N$ as functions of transverse field $h$ for (a) $J=0$ and (b) $J=0.5$. The dashed lines denote the locations of quantum critical points $h_c = 1 \pm J$ in the thermodynamical limit.}\label{fig:FS}
\end{figure}

Fig.~\ref{fig:FS}(a) shows the fidelity susceptibility per site, $\chi_F/N$, for $J=0$ across several system sizes $N$. As discussed in Sec.~\ref{sec:model}, the cluster–PM transition occurs at $h_c=1$. As the system size increases, the peak becomes progressively sharper, and its position approaches the exact critical point $h_c=1$. Nevertheless, finite-size effects remain clearly visible for the system sizes accessible to ED.

At $J=0.5$, the Hamiltonian in Eq.~\eqref{eq:Hami} hosts two Ising critical points at $h_c=1\pm J=0.5$ and $1.5$. As shown in Fig.~\ref{fig:FS}(b), two peaks emerge and progressively sharpen with increasing $N$ with their peak positions converging toward the exact critical fields indicated by the dashed lines. This behavior demonstrates that the fidelity susceptibility faithfully captures both quantum phase transitions, despite the finite-size effects that remain evident for small systems. These fidelity results provide a benchmark for assessing the performance of the Delta-VQE approach.

Although fidelity susceptibility provides a reliable method for identifying quantum critical points, its evaluation hinges on accurate calculations of ground states throughout the parameter space, and its critical signatures generally become well resolved only for sufficiently large system sizes to overcome finite-size effects.
In the following, we demonstrate that Delta-VQE can identify the same phase transitions using only shallow variational circuits and yields relatively accurate estimates of the critical points even for small system sizes.

\subsection{Delta-VQE: $J=0$}

\begin{figure}[t]
\includegraphics[width=0.95\linewidth]{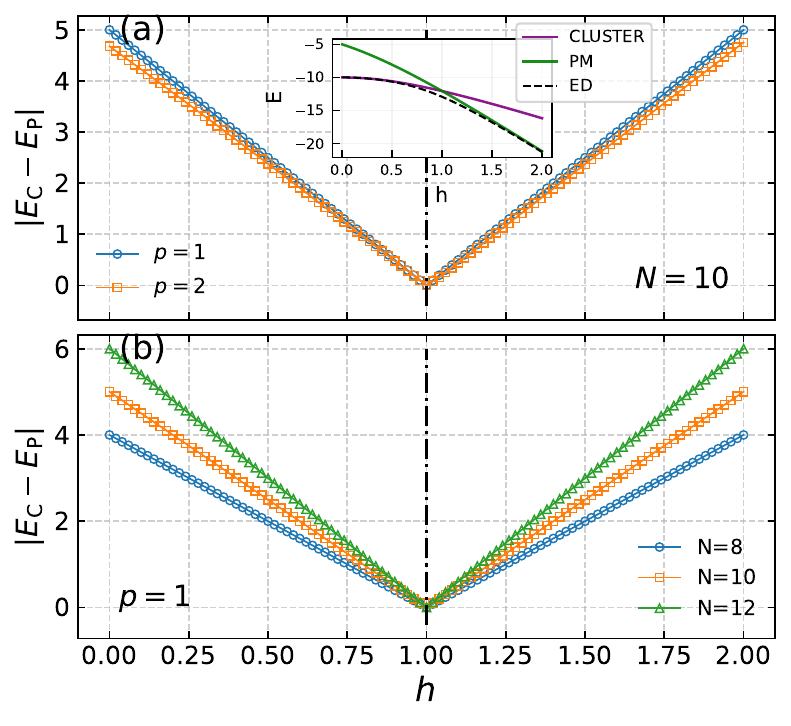}
\caption{
Delta-VQE results for the Ising coupling $J=0$, obtained with the variational ansatze $|\psi_\mathrm{P}(\bm{\theta})\rangle$ and $|\psi_\mathrm{C}(\bm{\theta})\rangle$. The dashed lines denote the location of quantum critical points $h_c = 1$ in the thermodynamical limit.
(a) Absolute energy difference, $\Delta E(h)=|E_\mathrm{C}-E_\mathrm{P}|$, as a function of the transverse field $h$ for a system of $N=10$ qubits with circuit depths $p=1$, 2. The inset compares the variational energies of the cluster and PM ansatze for $p=1$ with the exact ground-state energy evaluated by ED.
(b) Size dependence of $\Delta E(h)$ for $p=1$ with system sizes $N=8$, 10, and 12. Although the magnitude of $\Delta E(h)$ increases with $N$ away from the critical point, all curves exhibit a common minimum at the exact critical point $h_c=1$.
}
\label{fig:dVQE_J0}
\end{figure}

The above ED results provide a quantitative reference against which the performance of Delta-VQE will be assessed in the following subsections. We begin with the case $J=0$, where the system undergoes a cluster–PM transition at the critical field $h_c=1$. Notably, this critical point lies precisely on the self-dual line, as shown in Fig.~\ref{fig:phase_dia}.
To characterize this transition, we employ the variational ansatze $|\psi_\mathrm{P}(\bm{\theta})\rangle$ and $|\psi_\mathrm{C}(\bm{\theta})\rangle$ defined in Eqs.~\eqref{eq:psi_P} and~\eqref{eq:psi_C}. Fig.~\ref{fig:dVQE_J0}(a) displays the absolute energy difference $\Delta E(h)=|E_\mathrm{C}-E_\mathrm{P}|$ for a system of $N=10$ qubits with circuit depths $p=1$ and 2. Strikingly, even the shallowest circuit ($p=1$) produces a pronounced V-shaped profile with its minimum precisely at the exact critical point $h_c=1$. Increasing depth reduces the magnitude of $\Delta E(h)$, reflecting improved ground-state approximations, yet the minimum remains fixed. This demonstrates that Delta-VQE accurately identifies the critical point using only shallow variational circuits.

The inset of Fig.~\ref{fig:dVQE_J0}(a) compares the variational energies of the cluster and PM ansatze for $p=1$ with the exact ground-state energy obtained by ED. The optimized variational energies of the two ansatze closely follow the exact ground-state energy within their respective phases. As the transverse field approaches the critical point, however, the two variational energies approach one another, causing the energy difference to vanish. The equality of the two variational energies appears to imply that the two ansatze become nearly equivalent in describing the ground state near the critical point.

The size dependence of $\Delta E(h)$ for $J=0$ with $p=1$ is presented in Fig.~\ref{fig:dVQE_J0}(b). Away from the critical point, the magnitude of $\Delta E(h)$ increases with system size, whereas its minimum remains fixed at $h=1$. This behavior highlights an important aspect of Delta-VQE: even with shallow circuits, the minimum of $\Delta E(h)$ provides a stable and reliable indicator of the critical point despite finite-size effects. This feature of Delta-VQE has already been reported in Ref.~\cite{Shi_etal2023} for the transverse-field Ising model, the spin-anisotropic XZ model, and the cluster–Ising model. Notably, in all these examples, the quantum critical points coincide with the corresponding self-dual points, as is also the case here for $J=0$.

The apparent accuracy of Delta-VQE is particularly intriguing in view of the fidelity analysis shown in Fig.~\ref{fig:FS}(a), which exhibits pronounced finite-size effects for small systems. As we show in the next subsection, this coincidence between the critical point and the self-dual point provides an important clue to understanding the mechanism underlying the success of Delta-VQE.

\subsection{Delta-VQE: $J=0.5$ }\label{sec:J=0.5}

We next study the case of finite Ising coupling by setting $J=0.5$, for which the exact phase boundaries are located at $h_c=1\pm J=0.5$ and $1.5$. We begin by employing the same pair of ansatze, $|\psi_\mathrm{P}(\bm{\theta})\rangle$ and $|\psi_\mathrm{C}(\bm{\theta})\rangle$, and examine whether the diagnostic $\Delta E(h)=|E_\mathrm{C}-E_\mathrm{P}|$ continues to serve as a reliable indicator of these two critical points, in analogy with the fidelity susceptibility analysis shown in Fig.~\ref{fig:FS}(b).

\begin{figure}[t]
\includegraphics[width=0.95\linewidth]{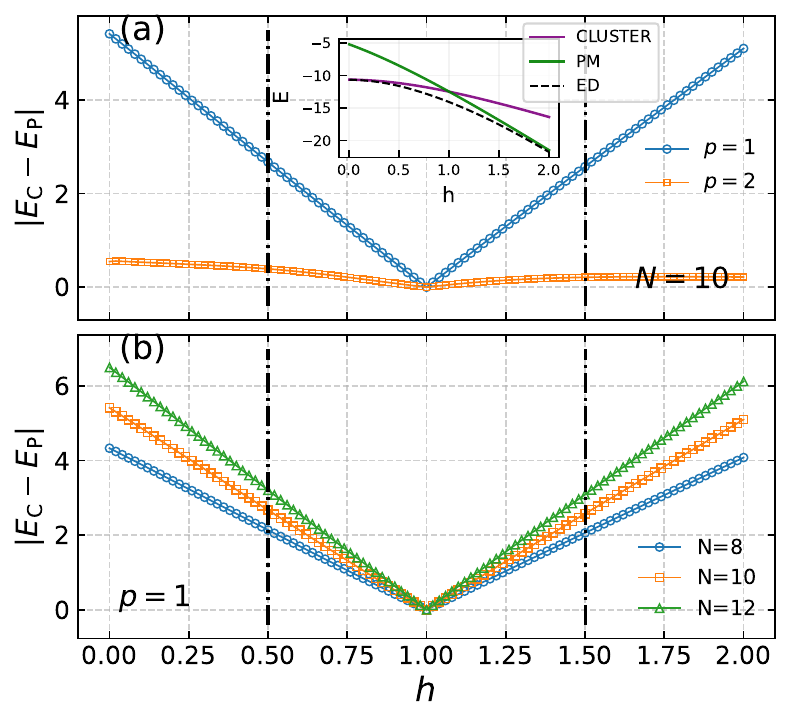}
\caption{
Same as Fig.~\ref{fig:dVQE_J0}, but for the Ising coupling $J=0.5$. The dashed lines denote the locations of quantum critical points $h_c = 1 \pm J=0.5$ and 1.5 in the thermodynamical limit. The minimum of $\Delta E(h)$ remains located at $h=1$ for circuit depths $p=1$, 2 and across all system sizes under consideration.
}
\label{fig:dVQE_J0.5_dual}
\end{figure}

The results are presented in Fig.~\ref{fig:dVQE_J0.5_dual}. Surprisingly, the minimum of $\Delta E(h)$ remains pinned at $h=1$ for both circuit depths shown in Fig.~\ref{fig:dVQE_J0.5_dual}(a), even though the true quantum critical points are located at $h_c=0.5$ and $1.5$. Increasing the circuit depth substantially reduces the overall magnitude of $\Delta E(h)$, but does not alter the location of its minimum. The finite-size dependence shown in Fig.~\ref{fig:dVQE_J0.5_dual}(b) leads to the same conclusion: although the energy difference increases with system size away from the minimum, the minimum itself stays at $h=1$. These observations reveal an important limitation of the Delta-VQE protocol.

Our findings admit a natural interpretation. As discussed at the end of Sec.~\ref{sec:construction}, the states $|\psi_\mathrm{C}(\bm{\theta})\rangle$ and $|\psi_\mathrm{P}(\bm{\theta})\rangle$ constitute a dual pair. Therefore, at the self-dual point the two variational energies are expected to coincide by symmetry, leading to $\Delta E(h=1)=0$. It is remarkable that this conclusion remains valid across all circuit depths $p$ and system sizes $N$, and does so independently of the Ising coupling $J$. That is, when a dual pair of ansatze is employed, the Delta-VQE protocol based on such a dual pair inevitably singles out the self-dual point. When this point happens to coincide with the genuine criticality, the identification becomes exact, persisting even for shallow circuits and small system sizes. This explains the results for the case $J=0$ and the self-dual models examined in Ref.~\cite{Shi_etal2023}. However, whenever the self-dual point deviates from the true critical points, Delta-VQE may incorrectly identify the quantum critical point.

This conclusion suggests that the effectiveness of Delta-VQE depends crucially on the choice of ansatze. Since the transition at $h_c=0.5$ separates the FM and cluster phases, it is most natural to adopt the associated ansatze, namely $|\psi_\mathrm{F}(\bm{\theta})\rangle$ and $|\psi_\mathrm{C}(\bm{\theta})\rangle$, as defined in Eqs.~\eqref{eq:psi_F} and~\eqref{eq:psi_C}, to capture this transition.

\begin{figure}[t]
\includegraphics[width=0.95\linewidth]{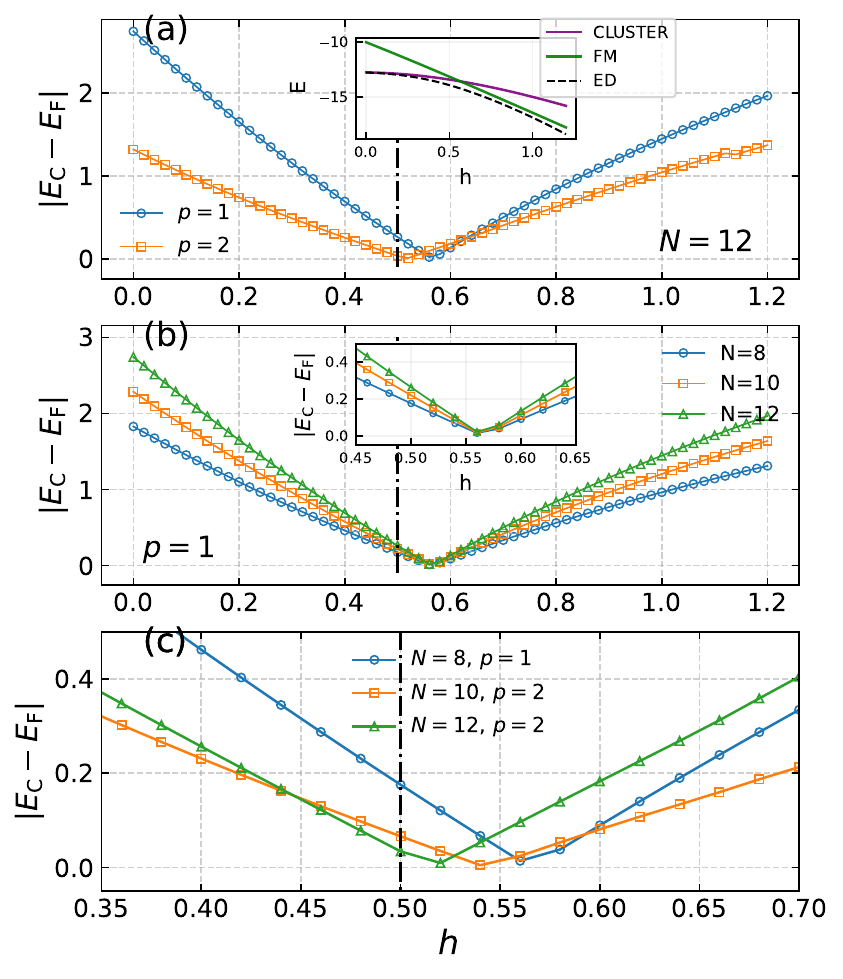}
\caption{
Delta-VQE results for the Ising coupling $J=0.5$, obtained with the variational ansatze $|\psi_\mathrm{F}(\bm{\theta})\rangle$ and $|\psi_\mathrm{C}(\bm{\theta})\rangle$. The dashed lines denote the location of quantum critical points $h_c = 0.5$ in the thermodynamical limit.
(a) Absolute energy difference, $\Delta E(h)=|E_\mathrm{C}-E_\mathrm{F}|$, as a function of the transverse field $h$ for a system of $N=12$ qubits with circuit depths $p=1$, 2. The inset compares the variational energies of the cluster and FM ansatze for $p=1$ with the exact ground-state energy evaluated by ED.
(b) Size dependence of $\Delta E(h)$ for $p=1$ with $N=8$, 10, and 12.
The inset provides a zoomed-in view.
(c) Size dependence of $\Delta E(h)$ under the condition $p=[N/4-\epsilon]$, with $[x]$ denoting the floor function and $\epsilon$ representing an infinitesimal positive offset.
}
\label{fig:dVQE_J0.5_FM}
\end{figure}

The corresponding results are shown in Fig.~\ref{fig:dVQE_J0.5_FM}. In contrast to Fig.~\ref{fig:dVQE_J0.5_dual}, the minimum of $\Delta E(h)=|E_\mathrm{C}-E_\mathrm{F}|$ is now shifted to the vicinity of the Ising critical point $h_c=0.5$. This demonstrates that the transition can indeed be identified once the correct pair of ansatze is employed. However, unlike the $J=0$ case, the minimum is not located precisely at the exact critical point. A similar observation has been reported for the XY–Ising model~\cite{Shi_etal2023}.
As illustrated in Fig.~\ref{fig:dVQE_J0.5_dual}(a), we further find that increasing the circuit depth to $p=2$ drives the minimum closer to the true critical point, thereby refining the estimate of $h_c$.

It is important to note that simply increasing system size at fixed depth does not necessarily yield further improvement in the estimate of $h_c$. This is because the expressivity of the variational circuit is governed primarily by its depth, and thus the expressibility of the trial wavefunction, is governed primarily by circuit depth. Indeed, for the HVA it has been established that the minimum depth required to prepare the ground state  scales linearly with system size~\cite{Ho-Hsieh2019,Wiersema_etal2020}. This perspective is supported by our results in Fig.~\ref{fig:dVQE_J0.5_FM}(b): although the magnitude of $\Delta E(h)$ increases with system size away from the transition, the minimum for circuits of fixed depth remains nearly size-independent and stays close to $h\approx0.56$.

Previous VQE studies have shown that the circuit depth $p$ must scale with the system size $N$ in order to faithfully reproduce thermodynamic-limit behavior~\cite{Bravo-Prieto_etal2020,Dreyer_etal2021,Jobst_etal2022,Schneider_etal2026}. Nevertheless, the precise scaling relation between $p$ and $N$ remains an open question and appears to depend sensitively on the choice of ansatz, as demonstrated in Ref.~\cite{Schneider_etal2026}. Within the Delta-VQE framework, the situation is further complicated because the energy difference $\Delta E$ is evaluated from two independently optimized variational states initialized from distinct ansatze and reference states. As a result, the extrapolation to the thermodynamic limit may depend not only on the system size and circuit depth, but also on the choice of ansatze and reference states. Since no established finite-size scaling framework currently exists to quantify these additional dependencies, a fully systematic extrapolation is not yet available. We therefore adopt the finite-depth scaling prescription proposed in Ref.~\cite{Dreyer_etal2021}, which showed that, once the condition $N\gtrsim 4p$ is satisfied, the finite-size effects will arise from the finite circuit depth $p$ rather than the finite system size $N$. Guided by this result, we examine the circuit-depth dependence of $\Delta E(h)$ under the condition $p \lesssim N/4$, thereby ensuring consistency with the finite-depth scaling relation. The corresponding results are shown in Fig.~\ref{fig:dVQE_J0.5_FM}(c).
Our data reveal that, as the system size increases, the estimated critical point systematically approaches the exact value $h_c=0.5$, indicating that finite-size corrections diminish with increasing $N$.

Importantly, for the same system size, Delta-VQE provides a significantly more accurate estimate of the critical point than the fidelity susceptibility peak [shown in Fig.~\ref{fig:FS}(b)]. This observation indicates that, despite its general applicability, fidelity susceptibility analysis yields a less reliable estimate of the critical point due to pronounced finite-size effects. By contrast, Delta-VQE achieves a markedly improved determination, particularly when only small system sizes are accessible.

The benefit of the Delta-VQE protocol can be understood as follows. At the critical field $h_c$, we proceed by decomposing the optimized ansatze, $|\psi_1(\bm{\theta}_1^*)\rangle$ and $|\psi_2(\bm{\theta}_2^*)\rangle$, into the exact eigenstates $\{|n\rangle\}$ ($n=0, 1, 2, \dots$) defined at this critical point:
\begin{align}
|\psi_1(\bm{\theta}_1^*)\rangle &= \sum_n a_n |n\rangle \;,\\
|\psi_2(\bm{\theta}_2^*)\rangle &= \sum_n b_n |n\rangle \;,
\end{align}
where $a_n$ and $b_n$ are the corresponding expansion coefficients. Consequently, the energy difference $\Delta E(h_c,J) = |E_1(h_c,J) - E_2(h_c,J)|$ defined in Eq.~\eqref{eq:delta_E} can be expressed as:
\begin{equation}
\Delta E(h_c,J) = \sum_{n>0} \left( |a_n|^2 - |b_n|^2 \right) \Delta\epsilon_n \;,
\end{equation}
where $\Delta\epsilon_n \equiv \epsilon_n - \epsilon_0$ denotes the excitation energy with $\epsilon_n$ being the eigenenergy associated with the state $|n\rangle$. For well-optimized variational states, the spectral weights, $|a_n|^2$ and $|b_n|^2$, are predominantly concentrated in the low-lying energy levels. Thus, even if the optimized states $|\psi_1(\bm{\theta}_1^*)\rangle$ and $|\psi_2(\bm{\theta}_2^*)\rangle$ deviate from the exact ground state (yielding nonzero weights for excited states), $\Delta E(h_c,J)$ can nevertheless remain small as long as $|a_n|^2 \approx |b_n|^2$ for all low-lying levels.
Under these conditions, the finite-size effects inherent in $|a_n|^2$ and $|b_n|^2$ partially cancel, and as a result, the critical field determined by the Delta-VQE protocol exhibits significantly reduced finite-size dependence.

\begin{figure}[t]
\includegraphics[width=0.95\linewidth]{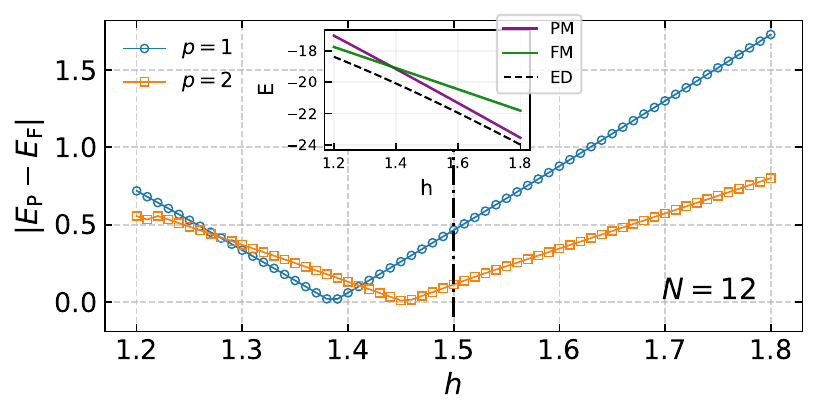}
\caption{
Delta-VQE results for the Ising coupling $J=0.5$, obtained with the variational ansatze
$|\psi_\mathrm{F}(\bm{\theta})\rangle$ and
$|\psi_\mathrm{P}(\bm{\theta})\rangle$.
The dashed line denotes the quantum critical point
$h_c=1.5$ in the thermodynamic limit.
Absolute energy difference,
$\Delta E(h)=|E_\mathrm{P}-E_\mathrm{F}|$,
as a function of the transverse field $h$
for a system of $N=12$ qubits with circuit depths
$p=1$ and 2.
The inset compares the variational energies of the FM and PM ansatze for $p=1$
with the exact ground-state energy obtained by ED.
}
\label{fig:dVQE_J0.5_FM_v2}
\end{figure}

Besides, in line with the preceding analysis, the FM–PM transition at $h_c=1.5$ can be resolved by adopting the ansatze corresponding to the two competing phases across the transition, namely $|\psi_\mathrm{F}(\bm{\theta})\rangle$ and $|\psi_\mathrm{P}(\bm{\theta})\rangle$, as specified in Eqs.~\eqref{eq:psi_F} and~\eqref{eq:psi_P}. The corresponding results are presented in Fig.~\ref{fig:dVQE_J0.5_FM_v2}.

Finally, to further support the general applicability of the Delta-VQE protocol, we investigate an extended transverse-field Ising model in App.~\ref{appendix:eTFIM}. The advantage of Delta-VQE in identifying critical points is once again confirmed.

\section{conclusions}\label{sec:conclusion}

In this work, we investigate the broad applicability and operational foundations of Delta-VQE, a resource-efficient quantum algorithm tailored to detect phase transitions with shallow variational circuits. Unlike approaches that rely on accurate ground-state preparation, Delta-VQE identifies quantum critical points by locating the minimum absolute energy difference between variational states optimized for competing phases of matter.

Using the generalized cluster-Ising model in Eq.~\eqref{eq:Hami} as an illustrative example, and using fidelity susceptibility calculated by ED as numerical benchmarks, our simulations reveal that when dual variational ansatze are employed, the energy difference $\Delta E$ inevitably vanishes at the self-dual point across all circuit depths $p$ and system sizes $N$. Consequently, when the true critical point does not coincide with the self-dual point, the Delta-VQE protocol based on a dual pair of ansätze always identifies the self-dual point rather than the genuine quantum critical point. However, when the ansatze are specifically chosen to represent the competing physical phases across the boundary, the minimum of $\Delta E$ accurately tracks the true Ising critical point with minor finite-size dependence. As discussed in Refs.~\cite{Bravo-Prieto_etal2020,Dreyer_etal2021,Jobst_etal2022,Schneider_etal2026}, a finite circuit depth $p$ introduces an emergent correlation length $\xi$, which grows algebraically with $p$. Since the characteristic width $|h-h_c|$ of the critical regime scales as $\xi^{-1/\nu}$, where $\nu$ denotes the correlation-length critical exponent, the critical window systematically narrows as the circuit depth increases. Consequently, the minimum of $\Delta E$ becomes increasingly localized near the critical point, and its position is expected to converge toward the thermodynamic critical value, resulting in progressively more accurate estimates of the critical point. Overall, our results clarify both the capabilities and limitations of Delta-VQE, emphasizing the crucial role of ansatz design in extending its general validity.

More generally, Delta-VQE may be conceived as a detector of level crossings. This suggests that it is particularly well suited to probing first-order quantum phase transitions characterized by ground-state crossings, provided that optimized variational circuits of modest depth can faithfully simulate the competing ground states.
In parallel, the equivariant VQE introduced in Ref.~\cite{Crognaletti_etal2025}, designed to locate the Berezinskii–Kosterlitz–Thouless transition through monitoring level crossings of symmetry-distinguished low-lying excited states, can be considered as a variant of Delta-VQE. This perspective opens a promising avenue for diagnosing transitions where traditional order parameters prove inadequate.
We anticipate that extending the Delta-VQE framework through physically motivated ansatz design will further broaden its reach to other classes of quantum criticality.

\begin{acknowledgments}
C.Y.H. gratefully acknowledges financial support from the National Science and Technology Council of Taiwan under Grant NSTC 114-2112-M-029-007 and NSTC 115-2112-M-029-003. M.-F.Y. was funded  by the National Science and Technology Council of Taiwan under Grant NSTC 114-2112-M-029-006, NSTC 115-2112-M-029-004 and NSTC 115-2112-M-029-007.
\end{acknowledgments}

\section*{DATA AVAILABILITY}
The data that support the findings of this article are not publicly available upon publication because it is not technically feasible and/or the cost of preparing, depositing, and hosting the data would be prohibitive within the terms of this research project. The data are available from the authors upon reasonable request.

\appendix
\section{Delta-VQE for extended transverse-field Ising model} \label{appendix:eTFIM}

In this appendix, we provide further numerical support for the effectiveness of Delta-VQE. To this end, we consider the transverse-field Ising model supplemented by a symmetry-preserving interaction that was previously inv estigated in Ref.~\cite{Jobst_etal2022}:
\begin{equation}\label{eq:eTFIM}
H=-\sum_{i} X_{i} X_{i+1} -h \sum_{i} Z_{i} -K \sum_{i} Z_{i} Z_{i+1} \, .
\end{equation}
Because the added term, $-K H_{ZZ} = -K \sum_{i} Z_{i}Z_{i+1}$, does not alter the underlying $\mathbb{Z}_2$ symmetry of the Ising model, the resulting quantum phase transition continues to belong to the Ising universality class. For the conventional transverse-field Ising model ($K=0$), the critical field $h_c=1$ can be obtained exactly through the Jordan-Wigner transformation to free fermions. For finite $K$, however, the model is no longer exactly solvable, and the critical field must be determined numerically. Here we focus on the case $K=0.5$, for which density-matrix renormalization-group (DMRG) calculations yield a critical field $h_c=0.295$~\cite{Jobst_etal2022}.

To capture the competition between the FM and PM phases, we employ the following phase-representative variational ansatze:
\begin{align}
&|\psi_\mathrm{F}(\bm{\theta})\rangle = \prod_{\ell=1}^{p} e^{i\,\theta_3^{(\ell)} H_{XX}}\, e^{i\,\theta_2^{(\ell)} H_Z}\, e^{i\,\theta_1^{(\ell)} H_{ZZ}}\, |\psi_\mathrm{F}\rangle \;, \label{eq:psi_F2} \\
&|\psi_\mathrm{P}(\bm{\theta})\rangle = \prod_{\ell=1}^{p} e^{i\,\theta_3^{(\ell)} H_Z}\, e^{i\,\theta_2^{(\ell)} H_{ZZ}}\, e^{i\,\theta_1^{(\ell)} H_{XX}}\, |\psi_\mathrm{P}\rangle \;. \label{eq:psi_P2}
\end{align}
The reference states $|\psi_\mathrm{F}\rangle$ and $|\psi_\mathrm{P}\rangle$ [see Eq.~\eqref{eq:ref_states}] are chosen as ground states of $-H_{XX}$ and $-H_Z$, respectively, thereby representing the FM and PM limits of the model.

\begin{figure}[t]
\includegraphics[width=0.95\linewidth]{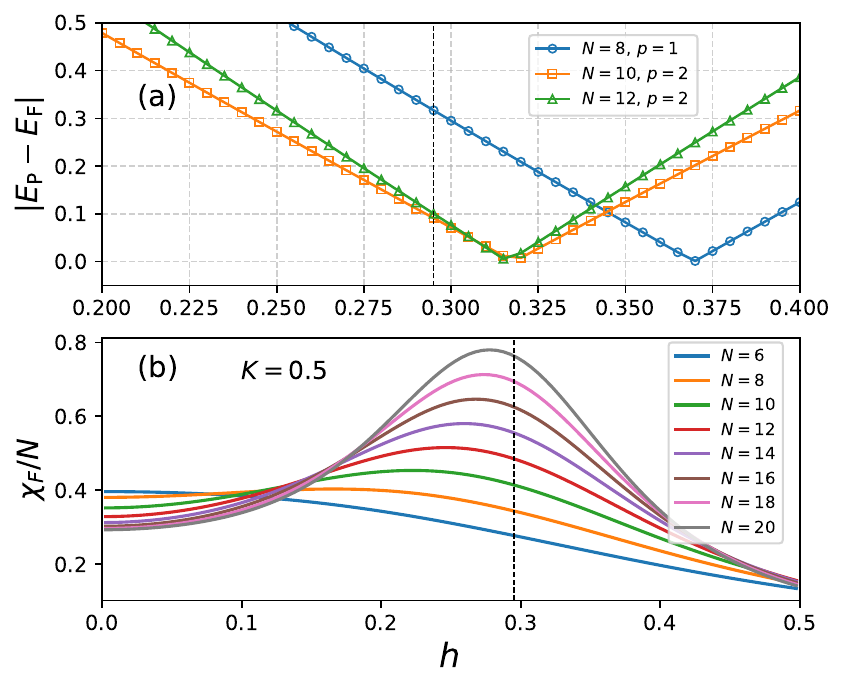}
\caption{Numerical results for the extended transverse-field Ising model in Eq.~\eqref{eq:eTFIM} with $K=0.5$.
(a) Size dependence of the absolute energy difference, $\Delta E(h)=|E_\mathrm{P}-E_\mathrm{F}|$, obtained from the phase-representative variational ansatze in Eqs.~\eqref{eq:psi_F2} and \eqref{eq:psi_P2}. The circuit depth is chosen according to $p=[N/4-\epsilon]$, where $[x]$ denotes the floor function and $\epsilon$ is an infinitesimal positive offset.
(b) Fidelity susceptibility per site, $\chi_F/N$, as a function of the transverse field $h$ with systems sizes up to $N=20$.
The vertical dashed lines in both panels indicate the critical field $h_c=0.295$ determined from DMRG calculations~\cite{Jobst_etal2022}.
}
\label{fig:dVQE_K0.5}
\end{figure}

The numerical results of Delta-VQE and the fidelity analysis for the present extended transverse-field Ising model are presented in Fig.~\ref{fig:dVQE_K0.5}. %
According to the same procedure described in Sec.~\ref{sec:J=0.5}, the data shown in Fig.~\ref{fig:dVQE_K0.5}(a) reveal that, as the system size increases, the estimated critical point, namely the location of the minimum of $\Delta E(h)$, can systematically approach the DMRG value $h_c=0.295$. This behavior reflects the diminishing influence of finite-size corrections with increasing $N$.
In particular, compared with the locations of the fidelity susceptibility peaks shown in Fig.~\ref{fig:dVQE_K0.5}(b), Delta-VQE provides a substantially more accurate estimate of the critical point for the same system size. This observation once again suggests that, despite its broad applicability, fidelity-susceptibility analysis is more susceptible to finite-size effects and therefore yields a less reliable estimate of the critical point. By contrast, Delta-VQE achieves a significantly improved determination of the critical point, especially when only small system sizes are available.

\section{Gate-noise simulations and zero-noise extrapolation} \label{appendix:noise}

As quantum hardware is highly susceptible to various sources of error, including decoherence, gate imperfections, and measurement inaccuracies, understanding the impact of noise is crucial for assessing the practical applicability of quantum algorithms. In this appendix, we investigate the sensitivity of the Delta-VQE protocol to realistic device noise through simulations of noisy quantum circuits.
We show that, by integrating standard error-mitigation techniques such as zero-noise extrapolation (ZNE)~\cite{Temme_etal2017,Li-Benjamin2017,Endo_etal2018,Giurgica-Tiron_etal2020}, the noiseless results can be successfully restored. These findings provide further evidence for the robustness of Delta-VQE in the NISQ regime.

To emulate noisy IBM Quantum processors, we first transpile the ansatz into the native gate set $\{\mathrm{CX}, X, \mathrm{SX}, R_Z\}$. Noise is applied only to the physical gates $(\mathrm{CX}, X, \mathrm{SX})$, while the phase-rotation gate $R_Z$ is assumed to be noiseless since it is implemented virtually. Reflecting the orders-of-magnitude difference in fidelity between single- and two-qubit operations, we adopt a single-qubit depolarizing error rate of $p_{1Q}=10^{-3}$ for the $X$ and $\mathrm{SX}$ gates, and a larger two-qubit depolarizing error rate of $p_{2Q}=10^{-2}$ for the $\mathrm{CX}$ gate. In addition to gate errors, readout errors are also present during the final measurement stage. In the present simulations, however, we neglect this source of noise, as highly effective mitigation techniques such as Twirled Readout Error Extinction (T-REX)~\cite{van_den_Berg_etal2022} are already available on current quantum devices.

\begin{figure}[t]
\includegraphics[width=0.98\linewidth]{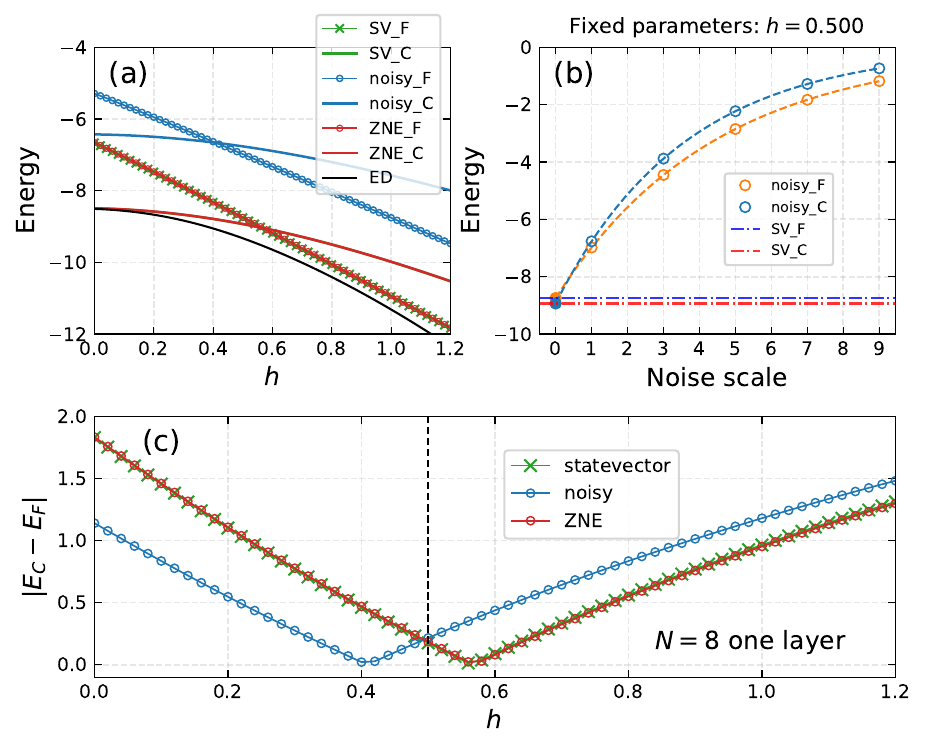}
\caption{Statevector (SV), noisy, and ZNE results for the generalized cluster-Ising model in Eq.~\eqref{eq:Hami} for $J=0.5$ with system size $N=8$ and circuit depth $p=1$.
(a) Variational energies obtained from SV (green), noisy (blue), and ZNE (red) simulations for the ansatze $|\psi_\mathrm{F}(\bm{\theta})\rangle$ (lines with circles) and $|\psi_\mathrm{C}(\bm{\theta})\rangle$ (solid lines without markers). The black curve indicates the exact ED ground-state energy for comparison.
(b) Exponential fits of the noisy expectation values $E(\lambda)$ as a function of the noise scale factor $\lambda\in\{1, 3, 5, 7, 9\}$ at $h = 0.5$. Dashed curves represent fits to the three-parameter exponential model $E(\lambda) = A e^{-\gamma\lambda} + C$. Solid symbols mark the extrapolated zero-noise values ($E_\mathrm{ZNE}$ at $\lambda = 0$), and horizontal dash-dotted lines indicate the ideal statevector results for reference.
(c) Absolute energy difference $\Delta E(h)=|E_\mathrm{C}-E_\mathrm{F}|$. The minimum in $\Delta E$ is shifted from $h_c = 0.40$ (noisy) to $h_c = 0.56$ (ZNE), closely matching the noiseless SV benchmark. The dashed vertical line marks the critical point $h_c = 0.5$ in the thermodynamic limit.
}
\label{fig:noise-zne}
\end{figure}

To mitigate the effects of gate-induced noise, the expectation values are post-processed using the ZNE technique. Specifically, we implement the local gate-folding protocol~\cite{Giurgica-Tiron_etal2020} to systematically amplify noise at odd-integer scale factors, $\lambda \in \{1, 3, 5, 7, 9\}$. The energy expectation values obtained from density-matrix simulations are then fitted to the three-parameter exponential ansatz $E(\lambda) = A e^{-\gamma \lambda} + C$~\cite{Endo_etal2018,Giurgica-Tiron_etal2020}, where $A$, $\gamma$, and $C$ serve as fitting parameters. Extrapolating this curve to the zero-noise limit ($\lambda = 0$) yields the error-mitigated ground-state energy $E_\mathrm{ZNE}$. This procedure allows us to evaluate the resilience of the Delta-VQE protocol against realistic gate noise and reliably recover noiseless performance.

As an illustrative example, we consider the FM-cluster transition in the generalized cluster-Ising model [Eq.~\eqref{eq:Hami}] at $J=0.5$ with system size $N=8$ and circuit depth $p=1$. Similar behavior is observed for larger system sizes and deeper circuits. The corresponding noiseless statevector simulations were presented in Sec.~\ref{sec:J=0.5}. In the noisy simulations, we employ the same phase-representative variational ansatze, $|\psi_\mathrm{F}(\bm{\theta})\rangle$ and $|\psi_\mathrm{C}(\bm{\theta})\rangle$, defined in Eqs.~\eqref{eq:psi_F} and~\eqref{eq:psi_C}. Owing to the stochastic nature of noise, which renders gradient estimation unreliable, the optimizer is replaced by the gradient-free COBYLA algorithm.

The optimized noisy energies $E(\lambda=1)$ and the corresponding error-mitigated values $E_{\mathrm{ZNE}}$ are plotted as functions of the transverse field $h$ in Fig.~\ref{fig:noise-zne}(a). For reference, the noiseless statevector results and the exact ED ground-state energies are also shown. The noisy variational energies are found to consistently lie above the noiseless values, reflecting the admixture of higher-energy states induced by gate noise. Consequently, the crossing point of the two variational energies, and thus the location of the minimum energy difference $\Delta E_{\mathrm{noisy}}=|E_C(\lambda=1)-E_F(\lambda=1)|$ shown in Fig.~\ref{fig:noise-zne}(c), is shifted away from its noiseless counterpart. In contrast, the variational energies obtained after applying ZNE almost completely recover the noiseless results. As a consequence, the minimum of $\Delta E_{\mathrm{ZNE}}$ is restored to nearly the same location as that obtained from the ideal statevector simulations. This demonstrates that the Delta-VQE protocol remains robust against realistic gate noise when combined with standard error-mitigation techniques.

To validate the use of exponential fitting, the noisy expectation values $E(\lambda)$ at $h=0.5$ are displayed as a function of the noise scale factor $\lambda$ in Fig.~\ref{fig:noise-zne}(b). Excellent agreement with the fitted curves is observed. The extrapolated zero-noise values, $E_{\mathrm{ZNE}}$ at $\lambda=0$, are found to recover the corresponding ideal statevector results with high accuracy. We note that the variational energies in the statevector calculations and in the noisy simulations are optimized independently, often leading to substantially different optimal parameter sets. Hence, the observed agreement between $E_{\mathrm{ZNE}}$ and the noiseless variational energies is far from trivial, and demonstrates that the ZNE procedure can effectively recover the underlying noiseless physics.



\end{document}